\documentclass{jetpl}
\twocolumn

\lat

\usepackage{graphicx}
\usepackage[noadjust]{cite}

\usepackage{txfonts}
\graphicspath{ {./} }

\title{RF-coil with variable resonant frequency for multiheteronuclear ultra-high field MRI}

\rtitle{RF-coil for multiheteronuclear MRI}

\author{V.\,A.\,Ivanov$^{+}$\thanks{e-mail: viacheslav.ivanov@metalab.ifmo.ru},
A.\,A.\,Hurshkainen$^{+}$,
G.\,A.\,Solomakha$^{+}$,
M.\,A.\,Zubkov$^{+}$}

\address{$^{+}$Faculty of Physics and Engineering, ITMO University,
199034 Saint-Petersburg, Russia}

\rauthor{V.\,A.\,Ivanov et al.}

\sodauthor{Ivanov, Hurshkainen, Solomakha, Zubkov}

\abstract{Here we propose double-coil setup to allow high signal-to-noise ratio broad-range heteronuclear magnetic resonance imaging experiments: two independent coils, one of them tuned to  $^{1}$H frequency to perform anatomical $^{1}$H imaging, and another one, metamaterial-inspired coil, tuned to the X-nucleus frequency. In this work our goal was to design a broad-range X-nuclei coil to cover $^{2}$H, $^{11}$B, $^{13}$C, $^{23}$Na, $^{7}$Li and $^{31}$P nuclear magnetic resonance frequencies, and to combine it with $^{1}$H coil in one setup. The system was designed for 11.7 T scanner, i.e., with 76-203 MHz frequency tuning range for the X-nuclei and tuned to 500 MHz for the proton coil. X-nuclei coil operates via excitation of the fundamental eigenmode of an array of parallel non-magnetic wires. The excitation of the array is provided via non-resonant feeding loop inductively coupled to the resonator. In order to tune the X-coil over such a wide range, both structural capacitance and inductance of the coil were made variable; narrow range tuning of the $^{1}$H coil is achieved via conventional tuning-matching circuit. Here, the design principle and setup tunability were investigated in simulations and experimentally.}

\begin{document}

\maketitle
\section{Introduction}
Nuclear magnetic resonance (NMR) provides unique opportunities for modern clinical and preclinical biomedical studies. Magnetic resonance spectroscopy (MRS) and magnetic resonance imaging (MRI) are two main NMR-based techniques. MRI allows to form images of the anatomy and to observe the physiological processes in the body. Essential MRI system components are the magnet, and the gradient and the radiofrequency (RF) system, the essential part of the latter being the RF coils. The magnet produces strong constant magnetic field $B_0$ that gives rise to macroscopic nuclear spin magnetization in the sample. The gradient system is used to spatially resolve the source of the NMR signal and thus form an MR image. RF coils, being transmit-only, receive-only or transceiver coils, play an important role of antennas providing alternating RF magnetic field to excite spins in a subject and receiving back weak echo signals at the Larmor frequency. The Larmor frequency of a nucleus (which is the operational frequency of an MRI system) is determined by the nucleus gyromagnetic ratio $\gamma$ and the static magnetic field strength
\begin{equation}
\omega_0 = \gamma B_0
\end{equation}
\begin{figure*}[t]
\includegraphics[width=\textwidth]{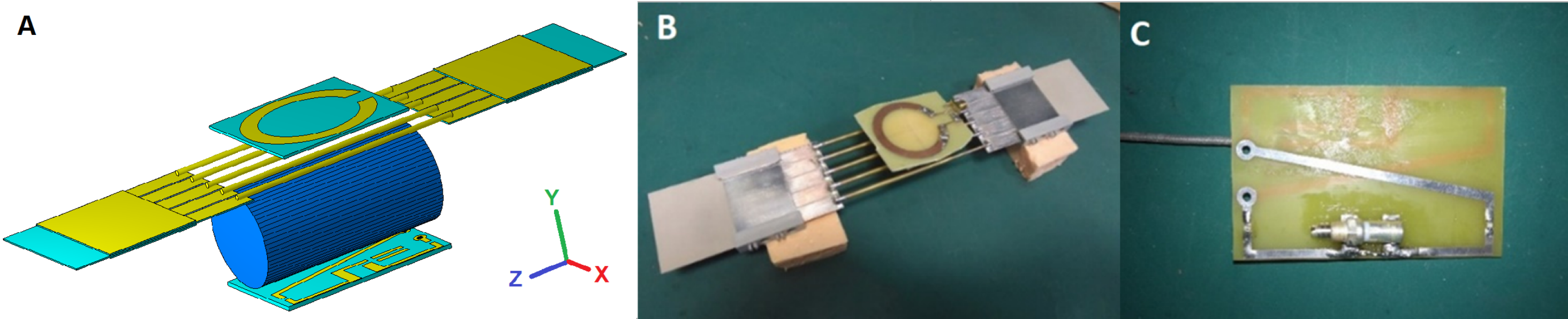}
\caption{Fig.1. A.~Numerical model of the multi-nuclei metamaterial-inspired coil. B.~Manufactured wire resonator with adjustable outer PCBs and inductively coupled feeding loop. C.~Manufactured butterfly coil with variable tuning/matching capacitor.}
\label{fig_1}
\end{figure*}
In clinical and research MRI, hydrogen nuclei (commonly referred to as protons) are most often used to generate a detectable RF signal. However, in addition to hydrogen, several other types of nuclei that provide a useful magnetic resonance response are present in living organisms~\cite{UFN_z}. Heteronuclear MR measurements, i.e., experiments using data from an additional non-proton nucleus, termed X-nucleus, such as deuterium ($^2$H), phosphorus ($^{31}$P), sodium ($^{23}$Na) or fluorine ($^{19}$F), are also of great value as they offer information which is complementary to that of proton MRI. For example, the information provided by $^{31}$P MRI or NMR spectroscopy can be used to observe changes in energy metabolism and intracellular pH~\cite{pol_j_rad,bachert,albers}, which can then be used for clinical studies in, for example, oncology~\cite{Lehnhardt,Novak,Ross}. The main issue with X-nuclei NMR signal detection is that the relative sensitivity (a function of gyromagnetic ratio) and the concentration of X-nuclei in organic tissue is much less than that of $^{1}$H. Thus in most cases X-nucleus images have a significantly lower signal-to-noise ratio (SNR) and consequently lower resolution. SNR in MRI largely depends on $B_0$, but also depends on the electromagnetic properties of RF coils. RF coils provide $B_1^+$  excitation field and $B_1^-$ reception field, having the same spatial patterns according to the principle of reciprocity. $B_1^+$ and $B_1^-$ are oppositely circularly polarized and both rotate in the plane orthogonal to $B_0$. The $B_1^+$ field serves to excite the spins in the studied sample or subject. It is usually desirable for this excitation to be uniformly distributed throughout the volume of interest, therefore the $B_1^+$ field produced by the RF coil should be homogeneous enough in this volume.  As the SNR or resolution of X-nuclei images is intrinsically low, reliable anatomical data from most X-nuclei images cannot be obtained and they have to be combined with the reference $^1$H images. In such cases, multichannel setups consisting of two or more coils tuned to the corresponding nuclei resonant frequencies are used to acquire X-nuclei and reference $^1$H images.

Whole-body small-animal metamaterial-inspired RF coils were successfully applied for {\it in-vivo} $^1$H mouse MRI~\cite{smallmouse_z} as well as dual-nuclei ($^{19}$F/$^1$H) scanning~\cite{smallmouse_h}. It has been shown~\cite{smallmouse_z} that this design allows higher SNR when compared to classic birdcage coil designs, and overcomes other drawbacks associated with conventional coils. A particular problem with both conventional coil design and the available metamaterial-inspired coils is the low tunability range of the X-nucleus coil. This manifests in the coils being produced for single- or dual-nuclei use only, thus severely limiting the applicability range of a particular coil. This problem escalates with the $B_0$ growth as the absolute frequency gap between the Larmor frequencies of two chosen nuclei grows. Therefore, it becomes quite challenging to design a wide band surface coil for ultra-high field scanners with a conventional matching-tuning circuit.

We propose a double-coil setup to allow broad range heteronuclear experiments: two independent coils, one of them tuned to $^1$H frequency to obtain anatomical reference images, and another, a metamaterial-inspired coil, tunable to a broad range of X-nuclei frequency. $^2$H and $^{31}$P frequencies were chosen as margin frequencies for the X-nucleus coil. The frequency gap between these two nuclei Larmor frequencies at 11.7 T is 127.7 MHz. Such NMR-active nuclei as $^7$Li, $^{11}$B, $^{13}$C, $^{23}$Na also fall into this range. We present here the results of numerical simulation and initial experimental implementation of such whole-body small-animal double-coil $^1$H-X system with a wide frequency tuning range for the X-nucleus to allow diverse heteronuclear experiments making use of data from multiple X-nuclei, i.e., multiheteronuclear imaging.
\section{Materials and methods}
Our coil system was designed to operate in 11.7 T small-animal scanner. The system incorporates 
two different RF-coils. One coil is designed to be placed above, the other below the imaged object. Both coils are aimed at producing $B_1^+$ magnetic field orthogonal to $B_0$ field inside the MR scanner bore. To facilitate the use of the coils they should not be coupled at their working frequencies. The way such decoupling was achieved will be demonstrated.

The first, multi-nuclei metamaterial-inspired coil operates via the excitation of hybridized eigenmodes of an array of parallel non-magnetic wires (schematic view is presented in Figure~\ref{fig_1},~A). The excitation both in transmission and reception regimes is provided via a non-resonant feeding loop inductively coupled to the resonator~\cite{smallmouse_h}. While multiple surface eigenmodes, all having different $B_1^+$ patterns, can be excited in such resonator type, the fundamental or the first eigenmode, having only one $B_1^+$ maximum in the scanning region~\cite{kosulnikov_fund}, has the most homogeneous $B_1^+$ distribution and the highest penetration depth into the subject. The fundamental mode frequency can be either the highest among other modes, or, as in our case, the lowest~\cite{smallmouse_h}. According to the range of nuclei desired to acquire the signal from ($^2$H-$^{31}$P), the first eigenmode of the resonator should be tunable in the frequency range 76-203 MHz.

As in the case with previously reported wire array coils~\cite{smallmouse_z,smallmouse_h}, the proposed coil requires no lumped elements to tune and match. It consists of 5 variable-length brass wires connecting two AD1000 PCBs (with $\epsilon$=10 and $\delta$=0.002 considered constant in our frequency range) with 5 metallization strips (or patches) on each side, the latter serving as one plate of the structural capacitance, that allows achieving the desired tunability range, while maintaining reasonable coil size. Each patch has the dimensions of 38$\times$9 mm, 1 mm gap between the patches. The opposing plates of the capacitors are formed by four AD1000 PCBs fully metallized on one side placed above and below PCBs with patches, thus forming an array of two parallel connected distributed capacitances. The substrate thickness of all PCBs is 0.5 mm. The distance between the wires is 10 mm. The wires have a 4 mm overlap with metal patches at both ends. Plastic holders are used to fix four outer PCBs with respect to the inner PCBs with patches. The coil is shown in Figure~\ref{fig_1},~B. The four outer boards can be moved relative to each other to change the overlap between the middle PCB with patches and outer PCBs, in this way varying the total capacity. The variation in the capacity attached to the wires changes the fundamental eigenmode frequency and thus allows the structure to be tuned to a particular frequency. On the other hand, changing the length of the wires is equal to changing distributed inductance of the structure, thus also allowing the resonant frequency of the coil to change. The coil therefore possesses double tuning capability. Matching of the coil is performed through selection of the optimal coupling between the resonator and the feeding loop which can be varied by modifying the distance between the loop and the resonator.

The second (lower) coil, the so-called butterfly loop coil~\cite{Kumar}, is tuned to 500 MHz (which corresponds to $^{1}$H resonance frequency at 11.7 T). It includes eight lumped capacitors, two of them variable, 2-10 pF to be used for tuning and matching, and six with fixed capacitance, 2.2 and 12 pF.  The coil is implemented as a double-sided FR-4 PCB (with $\epsilon$=4.4 and $\delta$=0.02, for our frequency range we considered them constant) with 65$\times$41 mm board dimensions and 1.5 mm thickness. The coil is shown in Figure~\ref{fig_1},~C. This coil configuration was chosen because the RF magnetic field it produces is predominantly directed orthogonally to the RF magnetic field produced by the X-nuclei coil which serves to reduce the mutual coil coupling.

The radiofrequency coil design was simulated in the commercial software CST Microwave Studio 2017 (Computer Simulation Technology GmbH, Germany) using the Frequency Domain solver. The latter solves a three-dimensional electrodynamics problem by the Finite Element Method. Simulations were performed in the presence of the RF shield model (i.e., copper tube with an inner diameter of 82 mm), which in MRI experiments represents the scanner bore, and the homogeneous cylindrical phantom (70 mm length, 18 mm radius) with dielectric permittivity of 50 and electric conductivity of 0.5 S/m imitating average tissue material properties of a small animal~\cite{Kraszewski}.  In simulations including all setup elements (i.e., the butterfly loop coil, metamaterial-inspired coil, shield, and phantom) the distance between the two coils was 42 mm. Due to the wide frequency range, simulation was performed for two frequency subranges centered around $^{1}$H and X-nuclei working frequencies. For the X-nucleus subrange simulations were conducted for 3 different settings: 76, 125.7, 203 MHz. The subrange width was 50 MHz for all central frequencies. Adaptive meshing was used for each subrange. 125.7 MHz frequency was considered significant because it is the $^{13}$C resonance frequency, and it is close to the center of the range. А 0.02 mm gap of air between the middle PCBs and the outer PCBs was added to make numerical simulation more realistic as it is impossible to provide a tight fit of the outer plates to the middle PCB evenly over the entire surface in the experiment. 
\begin{figure*}[t]
\includegraphics[width=\textwidth]{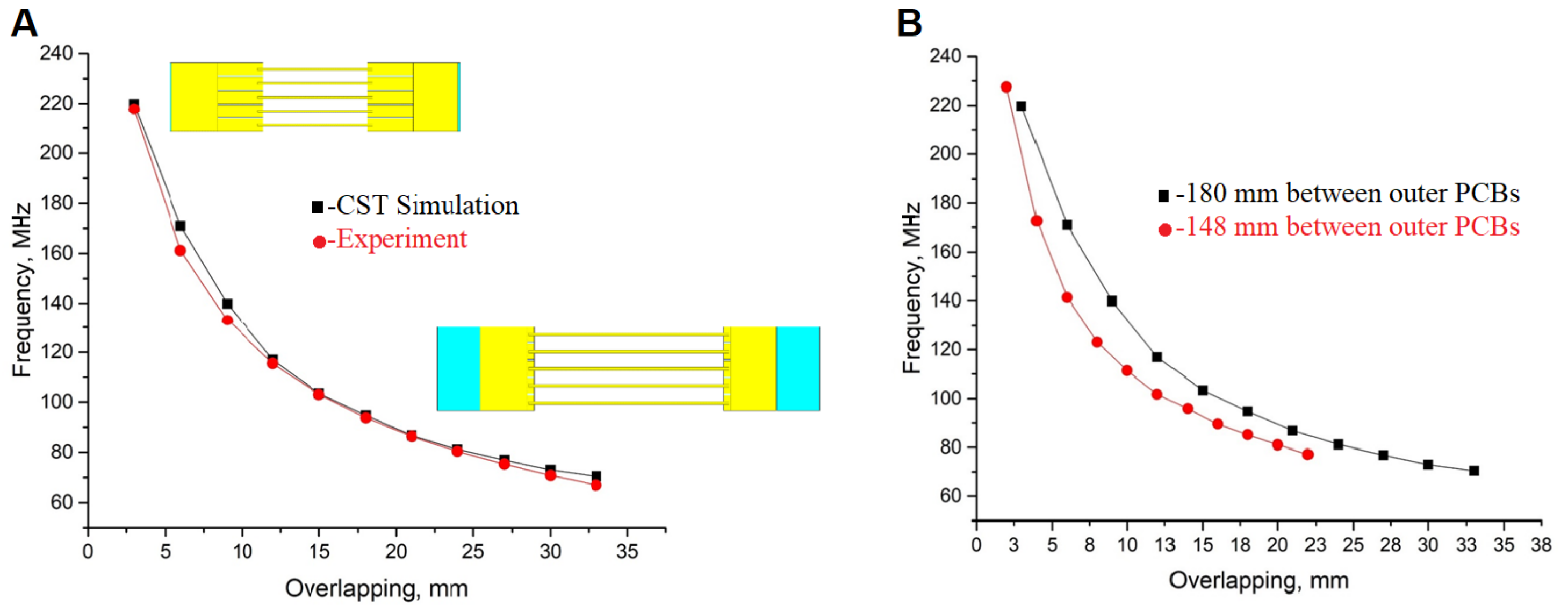}
\caption{Fig.2. A. Tuning range of the X-nuclei coil (simulation and experiment). B. Experimental curves of X-nuclei coil tunability for two different distances between outer PCBs (different system inductivity and different FoV).}
\label{fig_2}
\end{figure*}
\begin{figure}[t]
\includegraphics[scale = 0.35]{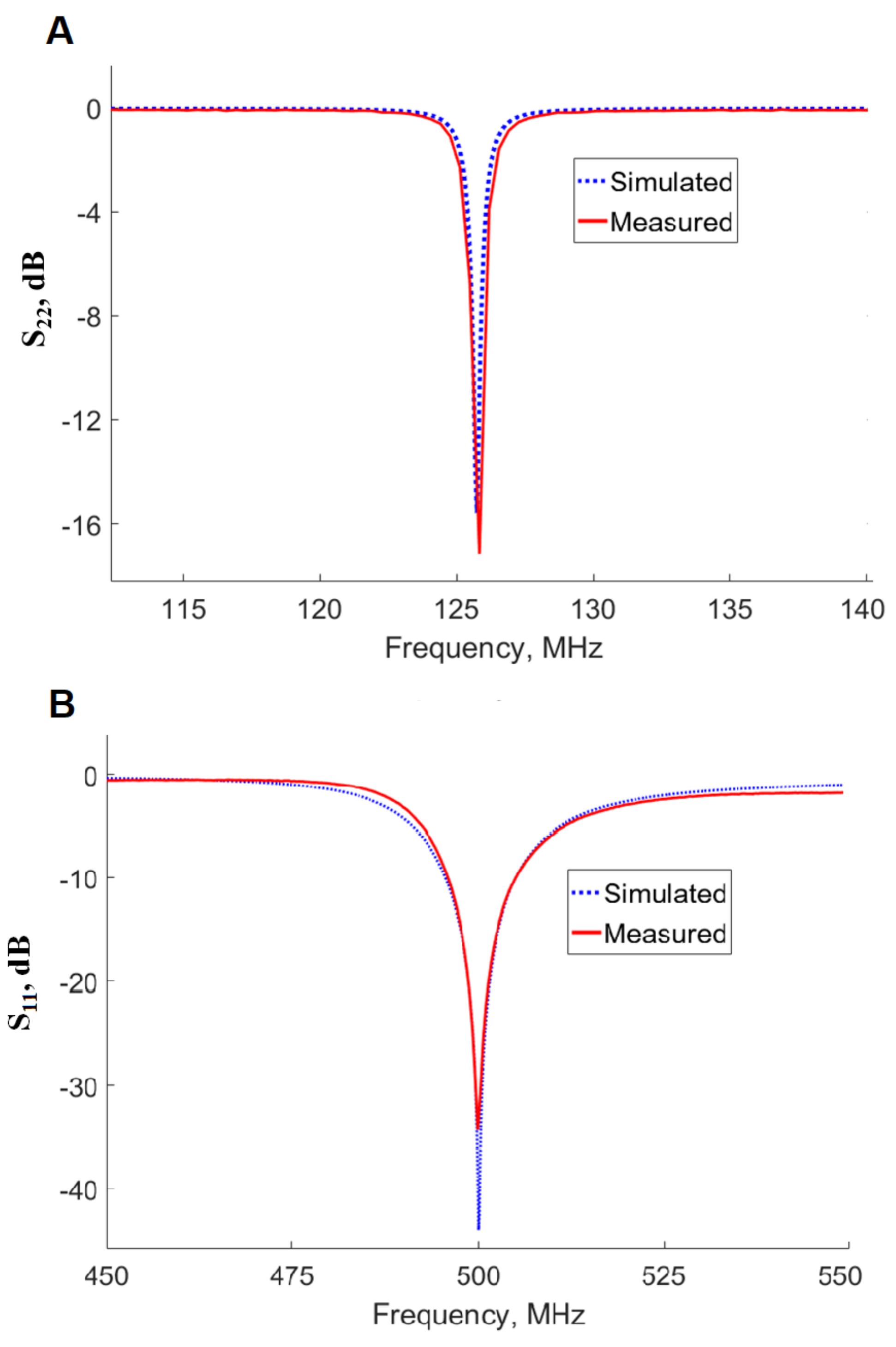}
\caption{Fig.3. Simulated and experimental S-parameters of the X-nuclei coil (A) and proton butterfly coil (B).}
\label{fig_3}
\end{figure}
\begin{figure*}[t]
\includegraphics[width=\textwidth]{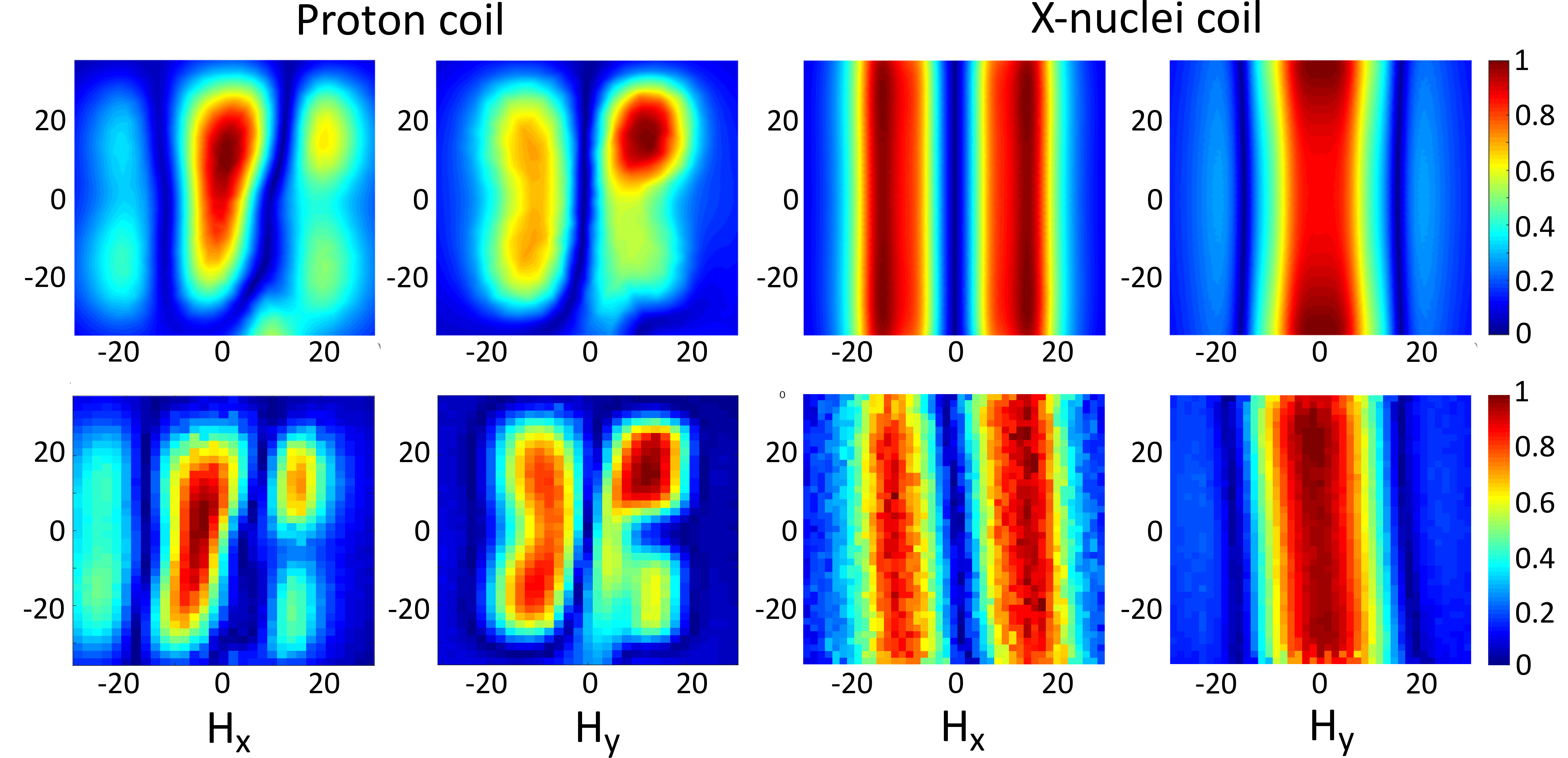}
\caption{Fig.4. Simulated (upper row) and experimental (lower row) magnetic field distribution. Normalized $H$-field components in XZ plane are presented. The dimensions are given in mm. Coordinate axes are indicated in Figure~1, A.}
\label{fig_4}
\end{figure*}
After determining the system parameters in the computer simulation, on-bench measurements were made. Necessary PCBs for both coils were printed by a specialized producer. A prototype of the multi-nuclei coil was assembled from PCBs and brass telescopic wires (the larger wire with an outer diameter of 3 mm, the smaller inner wire with an outer diameter of 2 mm). To prove that the prototype of the X-nuclei coil can be tuned to any frequency in the 76-203 MHz range, we gradually (with 3 mm steps) changed the overlap between the PCBs with patches and the outer PCBs (upper and lower PCBs from each side were moved synchronously), while simultaneously adjusting the length of the telescopic wires in order to keep the distance between the outer PCB at the two sides of the wires constant. The overlap between the PCBs with patches and the outer PCBs was altered from 3 to 33 mm, and the length of the telescopic wires from 78 to 138 mm respectively. At this point the measurement setup included the wire array and the feeding loop only, and the feeding loop $S_{11} $value was measured using Keysight PNA E8362C vector network analyzer (VNA). The $S_{11}$ parameter minimum position on the frequency axis, representing the wire array resonant frequency, was recorded for each overlap increment. Next, to be sure that X-nuclei and $^{1}$H coils were not interacting, $S_{11}$, $S_{12}$ and $S_{22}$ were measured with the $^{1}$H butterfly coil and feeding loop coil connected to port 1 and port 2 of the VNA respectively. The setup was placed into the RF shield and the coil positions were fixed with respect to the phantom and the shield. An RF shield much longer than the top (X-nuclei) coil was chosen, allowing the whole setup to be accommodated inside of it and simulate the conditions of a real MRI experiment. The measurements were repeated for three different tunings of the X-nuclei coil, as in the former numerical simulations.

\section{Results}
The simulations predicted a shift of the resonance frequency of the coils system when placed inside the RF-shield. This shift was about 6 MHz up for the 76 MHz tuning geometry, and about 20 MHz up for the 203 MHz tuning geometry. The same effect was observed during on-bench measurements. Simulation results and experimental measurements of the system tuning capability are presented in Figure~\ref{fig_2},~A. Tuning curves in Figure~\ref{fig_2},~B show that the proposed X-nuclei RF-coil can be tuned to any frequency in the target 76-203 MHz range for different fields of view (FoV).

The anticipated coil decoupling due to RF fields orthogonality was confirmed by calculated and measured $S_{12}$ parameters of the complete double-coil setup (Figure~\ref{fig_3}). While S-parameters were obtained for three X-nuclei coil geometries corresponding to three different system tuning frequencies (76, 125.7 and 203 MHz), for brevity, only the case of the middle frequency is presented in full here. $S_{11}$ value at $^{1}$H frequency is below -30 dB, and the $S_{22}$ value at the X-nuclei frequencies is below -12 dB for all three tested frequencies. $S_{12}$ value at $^{1}$H and X-nuclei frequencies in all three tuning configurations is below -30 dB.

The magnetic field distribution for both coils was obtained in simulations and experimentally. Since it was demonstrated that there is no interaction between the coils, these results were obtained independently for each coil and without the RF shield. Here we demonstrate the field components for the X-nuclei coil (map size is 80$\times$80 mm) and the proton coil (map size 70$\times$60 mm) (Figure 4). The asymmetry in the butterfly coil field maps is due to its asymmetric feeding by the coaxial cable which results in the cable antenna effect. The field maps of the X-nuclei coil confirm that we excited the fundamental eigenmode of the wire array. The field maps also provide the means to assess the FoV of the setup, the latter determined by the $B_1$ field magnitude distribution~\cite{Hoult_recip}. FoV of the X-nuclei coil can be seen to surpass that of the $^{1}$H coil in both X and Z directions. The maximum FoV of the current setup is therefore limited by the proton coil and is confined within the 50$\times$60 mm region.
\section{Summary and discussion}
The designed dual coil setup allows RF magnetic field excitation and acquiring signals from protons as well as from several biologically important nuclei at 11.7 T. The latter include $^{2}$H (76.753 MHz), which can be used, for example, to indicate tumor vascularity~\cite{Robinson}; $^{13}$C (125.721 MHz) with hyperpolarized $^{13}$C MRI promising to be a key to a better understanding of cancer metabolism~\cite{Kurhanewicz}; $^{23}$Na (132.256 MHz) which gives some important metabolic information such as tissue viability, through cell integrity and energy status~\cite{Madelin}; $^{129}$Xe (138.302 MHz), which in hyperpolarized state serves as a gaseous contrast agent for pulmonary MRI~\cite{Mugler} and $^{31}$P (202.404 MHz) being important in monitoring energy metabolism.

The S-parameter measurements have shown the butterfly coil to be uncoupled from the X-nuclei coil.
This allows independent tuning of the two coils, and thus provides a convenient way to obtain both the reference anatomical $^{1}$H images and functional X-nucleus images on the seleсted frequency.

Field distribution for both coils was shown to provide large enough FoV for whole-body small-animal imaging. However, the exact imaging volume  has to be determined with animal scanning in an actual MRI scanner.

An important detail of this design is that in the tuning process non-overlapped PCB parts effectively act as extensions of the connected wires. Therefore, in order to keep the FoV constant while switching between different X-nuclei, one needs to also keep the total length of telescopic wires and non-overlapped patches, i.e. the distance between outer PCBs, constant.

It has been previously demonstrated that, compared to the rectangular loop coils, $^{1}$H 300 MHz metamaterial-inspired coil shows a higher magnetic field generated per 1W of accepted power~\cite{smallmouse_z}. Since our adjustable tuning range coil operates in the same way, we expect higher SNR values for X-nuclei scanning, which in combination with its broad tuning range and large field of view presents a promising design for preclinical multinuclear MR-coils. 

Moreover, the possibility of seamlessly observing the NMR signal from a number of nuclei allows for better spatial localization of the imaging or localized spectroscopy data as well as removes the need to interfere with the studied animal when changing the scanner working frequency. This is expected to provide the ability to assign the obtained data to the same region and metabolic process. 

Briefly, the keypoints above show the presented coil to be a promising tool to allow previously unattainable practical muliheteronuclear MRI, thus allowing for better differentiation of pathologies, more detailed understanding of tumor metabolism or observing the functional aspects of the therapy progress, which will be shown in further, particularly {\it in-vivo}, experiments.

The experimental measurements in this work were supported by the Ministry of Education and Science of the Russian Federation (Zadanie No. 3.2465.2017/4.6). The numerical studies were funded by the Russian Science Foundation (Project 19-75-10104). This project has also received funding from the European Union's Horizon 2020 research and innovation programme under grant agreement No. 736937.

\end{document}